\documentclass[
aps,
prd,
reprint,
twocolumn,
nofootinbib,
superscriptaddress
]{revtex4-2}

\usepackage{amsmath,amssymb,amsthm}
\usepackage{physics}
\usepackage{bm}
\usepackage{graphicx}
\usepackage{hyperref}
\usepackage{booktabs}
\usepackage{xcolor}
\usepackage{float}
\usepackage{tikz}
\usetikzlibrary{decorations.pathmorphing, patterns}

\newtheorem{remark}{Remark}

\begin{document}

\title{
Born Geometry, Metaparticles, and an Effective Geometric Realization
of the Modular Black-Hole Remnant:
A Phase-Space Resolution of the Schwarzschild Singularity
}

\author{Paul-Robert Chouha} \email{paul.chouha@mcgill.ca} \affiliation{ Department of Physics, Ernest Rutherford Physics Building, McGill University, 3600 rue Universit\'e, Montr\'eal, Qu\'ebec H3A 2T8, Canada } \affiliation{ McGill University School of Continuing Studies, 680 Sherbrooke Street West, Montreal, Quebec H3A 3R1, Canada }

\date{\today}



\begin{abstract} Recent work on metaparticle black-hole thermodynamics suggested the existence of a finite evaporation endpoint characterized by a minimal horizon area, a maximal Hawking temperature, and a stable cold modular remnant. However, the corresponding spacetime geometry remained unknown. In this work we construct an effective geometric realization of the modular remnant within the framework of Born geometry, modular spacetime, and metaparticle dynamics. Starting from a doubled phase-space description in which spacetime coordinates are supplemented by dual coordinates, we derive a Born-doubled Schwarzschild geometry governed by a Born-invariant radial distance. The associated modular uncertainty relation prevents arbitrary localization at the phase-space origin, rendering the classical Schwarzschild singularity physically inaccessible and replacing it with a finite modular core. We then incorporate the metaparticle duality constraint, which induces an effective conformal deformation of the Born-geometric background and generates a nontrivial effective stress-energy tensor. The resulting interior region develops a finite anisotropic source characterized by a tension-dominated radial sector. Localized violations of the radial Null Energy Condition and the Strong Energy Condition arise naturally near the modular core, providing a mechanism through which the focusing assumptions underlying the Hawking--Penrose singularity theorems are evaded. These results establish an effective geometric counterpart of the modular remnant inferred from metaparticle black-hole thermodynamics and suggest that the Schwarzschild singularity is replaced by a finite Born-geometric phase-space structure whose ultraviolet behavior is governed by dual, non-geometric degrees of freedom.

\end{abstract}
\maketitle

\section{Introduction}
\label{Intro}

Black holes provide one of the most important arenas in which
gravitation, quantum theory, and thermodynamics intersect.
Despite its remarkable success in describing gravitational
phenomena over a vast range of scales, classical general
relativity predicts the formation of spacetime singularities
inside black holes. In the Schwarzschild solution, geodesic
evolution terminates at a curvature singularity located at
\(r=0\), where the classical spacetime description ceases to
be well defined. More generally, the Hawking--Penrose
singularity theorems establish that under suitable assumptions
concerning causal structure and the positivity properties of
matter, gravitational collapse inevitably leads to geodesic
incompleteness and singular behavior
\cite{Penrose1965,HawkingPenrose1970}.

The appearance of singularities is widely interpreted as signaling the breakdown of the classical spacetime description and the need for new physical principles at sufficiently short distances. Consequently, a large variety of mechanisms for singularity resolution have been proposed. Early examples include the regular black-hole geometries of Bardeen and Gliner \cite{Bardeen1968,Gliner1966}, while more recent approaches arise from loop quantum gravity, noncommutative geometry, asymptotic safety, limiting-curvature models, mimetic gravity, and string-theoretic constructions \cite{AshtekarBojowald2006,Nicolini2006, BonannoReuter2000,ChamseddineMukhanov2017, OlmoRubieraGarcia2015}. 

Within string theory, several distinct perspectives on the singularity problem have emerged. In the AdS/CFT correspondence, black holes in the bulk admit a dual description in terms of thermal states of the boundary conformal field theory, suggesting that the apparent bulk singularity may admit a well-defined microscopic interpretation beyond the classical spacetime description \cite{Maldacena1998,Witten1998}. A complementary viewpoint is provided by the fuzzball program of Mathur, in which black-hole microstates are associated with horizon-sized smooth geometries that replace the classical interior description of the black hole \cite{Mathur2005,Mathur2009}. Although the underlying mechanisms differ substantially, these approaches share the expectation that the classical spacetime description becomes incomplete in regimes where quantum-gravitational degrees of freedom play a dominant role.

Another aspect of string theory provides a particularly compelling motivation
for such a modification. One of its most remarkable
properties is T-duality, which relates large and small
distance scales and suggests that the distinction between
position and momentum variables is not fundamental
\cite{Amati1989}. These ideas are elevated to a foundational
principle in metastring theory and modular spacetime
\cite{FreidelLeighMinic2015,FreidelLeighMinic2016}. Within
this framework, spacetime is replaced by a doubled phase-space
geometry endowed with a symplectic structure, a polarization
metric, and a Born geometry. The fundamental excitations are
metaparticles, whose dynamics exhibit intrinsic UV/IR mixing
and encode a nontrivial interplay between geometric and dual
degrees of freedom
\cite{FreidelKowalskiLeighMinic2019}.

In a previous work
\cite{Chouha2026MetaparticleThermodynamics}, The thermodynamic analysis provided evidence for a finite evaporation endpoint characterized by a minimal horizon area, a maximal Hawking temperature, and a stable cold modular remnant. While these results strongly suggested the existence of a nontrivial ultraviolet completion of the Schwarzschild geometry, the corresponding spacetime realization remained unknown.

However, the existence of a thermodynamic remnant does not by
itself establish the regularity of the corresponding
spacetime. A minimal horizon area constrains the thermodynamic
evolution of a black hole, but it does not automatically imply
that spacetime trajectories avoid the classical Schwarzschild
singularity. Indeed, a finite horizon area and a regular
spacetime geometry are logically distinct notions. The
central question left unanswered by the thermodynamic analysis
is therefore whether the modular remnant admits a concrete
geometric realization and, if so, how the corresponding
spacetime avoids the singularity predicted by classical
general relativity. Equally important is the question of
whether the distinctive thermodynamic behavior of the
geometric and dual entropy sectors admits a direct geometric
interpretation within the underlying Born-geometric
phase-space structure.

The purpose of the present work is to address precisely these
questions. The construction proceeds in two stages. First,
starting from the Born-geometric structure underlying modular
spacetime \cite{FreidelLeighMinic2014, FreidelLeighMinic2015, FreidelLeighMinic2016}, we construct an effective  doubled Schwarzschild geometry in
which the ordinary radial coordinate is replaced by the
Born-invariant combination

\begin{equation}
\mathcal R^2
=
r^2
+
\lambda^4\tilde r^2.
\end{equation}

The resulting effective geometry possesses a finite modular core and
removes access to the classical phase-space origin through a
modular uncertainty relation. Second, we incorporate the
metaparticle duality constraint, which induces a conformal
deformation of the Born-doubled geometry. This generates a
nonvanishing Einstein tensor and an associated effective
stress-energy tensor whose interior behavior is characterized
by a finite tension-dominated anisotropic source.

We show that localized violations of the radial Null Energy
Condition and the Strong Energy Condition arise naturally
near the modular core, thereby providing a concrete mechanism
through which the focusing assumptions entering the
Hawking--Penrose singularity theorems are evaded. The
resulting spacetime possesses finite curvature invariants
throughout the physically admissible region, eliminates
physical geodesic incompleteness, and replaces the
Schwarzschild singularity with a finite modular core.

Taken together, these results provide the effective
Born-geometric realization of the thermodynamic remnant
proposed in \cite{Chouha2026MetaparticleThermodynamics}.
More generally, they suggest that the classical spacetime
description may represent an infrared manifestation of a
deeper Born-geometric phase-space structure. Within this
picture, the modular remnant is interpreted not merely as the
endpoint of black-hole evaporation, but as the onset of a
regime in which the dual, non-geometric sector becomes
dominant. As we shall show, this transition follows directly
from the Born-geometric structure itself: near the modular
core the geometric contribution to the invariant radius
becomes subdominant, while the dual sector remains finite as
a consequence of the modular uncertainty relation.

This paper is organized as follows. In
Section~\ref{Section II} we review the essential ingredients
of Born geometry, modular spacetime, and metaparticle
dynamics. In Section~\ref{Sec: III} we derive the
Born-invariant radial coordinate and construct the
Born-doubled Schwarzschild geometry. The regularity
properties of the resulting spacetime, including curvature
invariants, geodesic structure, and the modular core, are
investigated in Section~\ref{Sec: IV}. In
Section~\ref{Sec: V} we incorporate the metaparticle duality
constraint and derive the resulting effective conformal
geometry together with its associated effective source.
Section~\ref{Sec:VI} analyzes the effective stress-energy
tensor, the corresponding energy conditions, and the
mechanism through which the Hawking--Penrose focusing
assumptions are evaded. The connection between the present
geometric construction and the thermodynamic remnant of
Ref.~\cite{Chouha2026MetaparticleThermodynamics} is developed
in Section~\ref{Sec:VII}. We conclude in
Section~\ref{Sec:VIII} with a discussion of the broader
implications of these results and possible extensions to
generalized uncertainty principles, black-hole
phenomenology, and modular cosmology.
\section{Born Geometry, Modular Spacetime, and Metaparticles}\label{Section II}

\subsection{Modular Spacetime}\label{SubSec:Modular ST}

A fundamental lesson emerging from string theory is that the
conventional distinction between position and momentum variables
may not be fundamental. This insight is most clearly manifested
through T-duality, which relates large and small distance scales
and suggests the existence of a more symmetric description in
which spacetime and momentum space are treated on an equal footing
\cite{Amati1989}.

These ideas are elevated to a foundational principle in metastring
theory and modular spacetime
\cite{FreidelLeighMinic2015,FreidelLeighMinic2016}. Rather than
regarding spacetime as the primary arena of physical processes,
the fundamental description is formulated on a doubled phase space
whose coordinates are

\begin{equation}
\mathbb{X}^M
=
(x^\mu,\tilde{x}^{\rm phys}_\mu),
\end{equation}

where $x^\mu$ denote the usual spacetime coordinates and
$\tilde{x}^{\rm phys}_\mu$ represent their dual counterparts.

The doubled phase space is characterized by a fundamental
modular length scale $\lambda$, which plays a role analogous to
the string length in conventional string theory. The parameter
$\lambda$ sets the scale at which the geometric and dual sectors
become intrinsically coupled and determines the onset of
Born-geometric effects. In the infrared limit
$\lambda\rightarrow0$, the influence of the dual coordinates is
suppressed and ordinary spacetime is recovered.

\begin{remark}[Notation for the Dual Coordinates] \label{Rmk:r_phys} In the underlying metaparticle formulation, the fundamental dual coordinates, denoted by $\tilde x_\mu^{\rm phys}$, carry dimensions of length, \[ [\tilde x_\mu^{\rm phys}] = L. \] For the effective Born-geometric construction developed in Secs.~III and IV, it is convenient to introduce the rescaled coordinates \begin{equation} \tilde x_\mu = \frac{\tilde x_\mu^{\rm phys}}{\lambda^2}, \label{eq:phys_dual_coord} \end{equation} where $\lambda$ denotes the fundamental modular length scale. The rescaled coordinates therefore satisfy \[ [\tilde x_\mu]=L^{-1}. \] This rescaling serves two purposes. First, it ensures dimensional consistency between the geometric and dual sectors. Second, and more importantly, it makes the fundamental Born-geometric scale $\lambda$ appear explicitly in the effective line element and in the associated Born-invariant distance measure. In this way, the relative contribution of the dual sector is controlled directly by the modular scale that characterizes the underlying doubled geometry. Unless otherwise stated, the symbol $\tilde x_\mu$ will refer to the rescaled coordinates throughout the remainder of the present section. For spherically symmetric configurations, \[ \tilde r = \frac{\tilde r_{\rm phys}}{\lambda^2}, \] so that $\lambda^2\tilde r$ has dimensions of length and can be combined naturally with the spacetime radius $r$ in Born-invariant geometric quantities.  \end{remark}

The doubled formulation is equipped with a symplectic structure,
a neutral polarization metric, and a compatible generalized metric.
Together these structures define the modular phase space underlying
the metastring framework and provide a unified geometric description
of spacetime and its dual degrees of freedom.

An important consequence of the doubled formulation follows from
its underlying symplectic structure. The metastring and
metaparticle phase spaces are endowed with a nontrivial Poisson
algebra \cite{FreidelLeighMinic2015,FreidelLeighMinic2016,
FreidelKowalskiLeighMinic2019}\footnote{
A detailed discussion of the generalized Poisson structure,
its relation to Born geometry, and its implications for
Generalized Extended Uncertainty Principles (GEUPs) will be
presented elsewhere \cite{ChouhaGEUP}.
}

\begin{equation}
\{x^\mu,\tilde x_\nu\}_{\rm phys}
=
\pi\lambda^2
\delta^\mu{}_\nu ,
\label{eq:poisson_modular_intro}
\end{equation}

which directly couples the geometric and dual coordinates.
Upon quantization, Eq.~(\ref{eq:poisson_modular_intro})
implies

\begin{equation}
[\hat x^\mu,\hat{\tilde x}_\nu]_{\rm phys}
=
i\pi\lambda^2
\delta^\mu{}_\nu ,
\end{equation}

and therefore leads to the modular uncertainty relation

\begin{equation}
\Delta x^\mu\,
\Delta \tilde x^{\rm phys}_\mu
\gtrsim
\frac{\pi\lambda^2}{2}.
\label{eq:modular_uncertainty_intro}
\end{equation}

For spherically symmetric configurations this reduces to

\begin{equation}
\Delta r\,
\Delta \tilde r_{\rm phys}
\gtrsim
\frac{\pi\lambda^2}{2},
\end{equation}
where $\tilde{r}_{\rm phys}$ is defined in Remark~\ref{Rmk:r_phys}.

Unlike ordinary spacetime, modular spacetime therefore does not
permit simultaneous arbitrarily precise localization of geometric
and dual coordinates. In particular, the phase-space origin
$(r,\tilde r_{\rm phys})=(0,0)$ cannot be reached by physical states. As
will be shown below, this limitation lies at the heart of the
mechanism through which the Schwarzschild singularity is
eliminated.

\subsection{Born Geometry}\label{SubSec:Born Geometry}

The geometric structure underlying modular spacetime is known as
Born geometry \cite{FreidelLeighMinic2016,SvobodaRudolph2019}. A Born geometry is
characterized by three mutually compatible tensors:

\begin{equation}
\omega_{MN}
=
\begin{pmatrix}
0 & \delta^\mu{}_\nu \\
-\delta_\mu{}^\nu & 0
\end{pmatrix},
\end{equation}

\begin{equation}
\eta_{MN}
=
\begin{pmatrix}
0 & \delta^\mu{}_\nu \\
\delta_\mu{}^\nu & 0
\end{pmatrix},
\end{equation}

and

\begin{equation}
H_{MN}
=
\begin{pmatrix}
h_{\mu\nu} & 0\\
0 & h^{\mu\nu}
\end{pmatrix},
\end{equation}

where $h_{\mu\nu}$ denotes the spacetime metric and
$h^{\mu\nu}$ its inverse.

The tensors $\omega_{MN}$, $\eta_{MN}$ and $H_{MN}$ encode,
respectively, the symplectic structure of phase space, the
neutral metric associated with T-duality, and the generalized
Born metric. Together they satisfy the compatibility conditions
defining a Born geometry and implement Born reciprocity, namely
the symmetry between spacetime and momentum-space variables.

The generalized metric defines the invariant line element

\begin{equation}
ds_B^2
=
H_{MN}
\,d\mathbb X^M d\mathbb X^N,
\label{eq:born_line_element_intro}
\end{equation}

which replaces the ordinary notion of spatial distance. Physical
observables must therefore be constructed from quantities that are
invariant under the Born-geometric structure of the doubled space.

The significance of Born geometry for gravitational physics stems
from the fact that singularity formation in general relativity is
typically associated with arbitrarily precise spacetime
localization. In a modular phase-space description, however, the
uncertainty relation \eqref{eq:modular_uncertainty_intro} prohibits
simultaneous localization in both geometric and dual directions.
This suggests that the classical notion of a spacetime singularity
may lose operational meaning within the Born-geometric framework.

\subsection{Metaparticle Dynamics}\label{SubSec:Metaparticle Dynamics}

The elementary excitations of modular spacetime are
metaparticles, which arise as the zero-mode sector of the
metastring \cite{FreidelLeighMinic_MetaParticle,
FreidelKowalskiLeighMinic2019}. Unlike ordinary relativistic
particles, metaparticles propagate on the doubled phase space and
are characterized by a doubled momentum vector

\begin{equation}
\mathbb P_A
=
(p_\mu,\tilde p^\mu),
\end{equation}

containing both geometric and dual momentum components.

The dynamics are governed by two first-class constraints. The first
generalizes the usual relativistic mass-shell condition,

\begin{equation}
\mathcal H
=
\frac12
\left(
p^2+\tilde p^{\,2}+2\mathfrak{m}^2
\right)\approx 0,
\label{eq:Hconstraint_intro}
\end{equation}
where $\mathfrak{m}$ is the metaparticle mass-scale;
while the second couples the geometric and dual momentum sectors,

\begin{equation}
\mathcal D
=
p_\mu\tilde p^\mu
-
\mu\approx 0.
\label{eq:Dconstraint_intro}
\end{equation}

The constraint $\mathcal H$ governs the generalized propagation of
the metaparticle on the doubled phase space, whereas $\mathcal D$
encodes a nontrivial relation between geometric and dual degrees of
freedom. The latter introduces an intrinsic UV/IR mixing modulated by the duality-parameter $\mu$ and is
responsible for many of the characteristic features of the
metaparticle framework.

In the present work the Born geometry determines the kinematical
structure of the doubled black-hole spacetime, while the
metaparticle constraint $\mathcal D$ generates an effective
backreaction that modifies the corresponding gravitational
description. As we shall see, this induced deformation gives rise
to a nontrivial effective stress-energy tensor and plays an
essential role in the dynamical avoidance of singularity formation.
\section{Born-Doubled Schwarzschild Geometry}\label{Sec: III}

The central objective of this section is to construct the
spherically symmetric vacuum geometry associated with the
Born-geometric phase space introduced in the previous section.
Rather than postulating a modified radial coordinate, we derive
the appropriate radial variable directly from the Born metric and
then determine the corresponding vacuum solution.

\subsection{The Born-Invariant Radial Coordinate}\label{SubSec: Born-Inv Radial-coord}

The fundamental geometric object of modular spacetime is the
Born metric

\begin{equation}
ds_B^2
=
H_{MN}\,
d\mathbb X^M d\mathbb X^N,
\label{eq:BornLineElement}
\end{equation}

defined on the doubled phase space

\begin{equation}
\mathbb X^M
=
(x^\mu,\tilde x^{\rm phys}_\mu).
\end{equation}

For static and spherically symmetric configurations, the
spacetime and dual sectors admit independent radial coordinates
$r$ and $\tilde r$. 

As we show in detail in Appendix~\ref{Appendix:A} and restricting the Born line element to the
radial sector yields

\begin{equation}
ds_{B,\mathrm{rad}}^2
=
dr^2
+
\lambda^4 d\tilde r^2 ,
\label{eq:radialbornmetric}
\end{equation}

where the factor $\lambda^4$ guarantees dimensional consistency
between the geometric and dual coordinates.

Equation (\ref{eq:radialbornmetric}) identifies the natural
Born-invariant radial distance as

\begin{equation}
d\mathcal R^2
=
dr^2
+
\lambda^4 d\tilde r^2 .
\end{equation}

Integrating this relation immediately gives

\begin{equation}
\boxed{
\mathcal R^2
=
r^2
+
\lambda^4 \tilde r^2 .
\label{eq:BornInvariantRadius}
}
\end{equation}

The quantity $\mathcal R$ is therefore not introduced as an
additional assumption but follows directly from the radial
reduction of the Born metric{\footnote{At the level of fundamental Born geometry, the coordinates $r$ and $\tilde r$ remain independent coordinates of the doubled phase space. The emergence of the effective radial coordinate $\mathcal R$ should therefore not be interpreted as a kinematical identification of the geometric and dual sectors. Rather, the metaparticle duality constraint \( \mathcal D=p_\mu\tilde p^\mu-\mu=0 \) couples the two sectors dynamically and restricts the physical dynamics to the constraint surface \( \Sigma_\mu=\{(p_\mu,\tilde p^\mu)\,|\,p_\mu\tilde p^\mu=\mu\} \) within the doubled phase space. The physically admissible trajectories therefore belong to a distinguished class selected by \(\Sigma_\mu\), for which the Born-invariant radius provides a natural effective description. }}
Unlike the ordinary Schwarzschild radius, $\mathcal R$
incorporates both geometric and dual information and provides
the appropriate radial variable for any spherically symmetric
Born-geometric spacetime.

Furthermore, Eq.~(\ref{eq:BornInvariantRadius}) must be viewed
together with the modular uncertainty relation

\begin{equation}
\Delta r\,\Delta\tilde r
\gtrsim
\frac{\pi}{2},
\end{equation}

which prevents simultaneous localization of the geometric and
dual coordinates. Consequently, the phase-space origin is not a
physical point of the theory, suggesting that singularity
formation may be fundamentally altered in the Born-geometric
framework.

\subsection{Metric Ansatz}\label{SubSec:Metric-Ansatz}

In the absence of a generally accepted gravitational action principle for Born geometry, we adopt an effective spacetime description in which the Born-invariant coordinate $\mathcal R$ is treated as the radial variable appearing in a static and spherically symmetric metric ansatz. The resulting construction should therefore be interpreted as an effective geometric realization of the underlying Born-geometric phase-space structure rather than as a fundamental gravitational theory defined directly on the doubled space. Motivated by the Schwarzschild solution and by the requirement of asymptotic agreement with ordinary general relativity, we consider the metric ansatz

\begin{equation}
ds^2
=
-f(\mathcal R)\,dt^2
+
\frac{d\mathcal R^2}{f(\mathcal R)}
+
\mathcal R^2 d\Omega^2 ,
\label{eq:BornAnsatz}
\end{equation}

where

\begin{equation}
d\Omega^2
=
d\theta^2
+
\sin^2\theta\, d\phi^2
\end{equation}

is the metric on the unit two-sphere.

The form (\ref{eq:BornAnsatz}) preserves the standard
Schwarzschild structure while replacing the ordinary radial
coordinate by the Born-invariant quantity
(\ref{eq:BornInvariantRadius}). As a result, spherical symmetry
is formulated directly in the doubled phase-space geometry.

The unknown function $f(\mathcal R)$ is determined by the vacuum
field equations.

\subsection{Vacuum Solution}\label{SubSec:Vacuum_Solutions}

Within the effective spacetime description introduced above, we determine the metric function $f(\mathcal R)$ by imposing the ordinary vacuum Einstein equations on the effective metric (\ref{eq:BornAnsatz}),

\begin{equation}
G_{\mu\nu}=0,
\end{equation}

yields the nonvanishing Einstein tensor components

\begin{equation}
G^{t}{}_{t}
=
G^{\mathcal R}{}_{\mathcal R}
=
-\frac{1}{\mathcal R^2}
\left[
1-f(\mathcal R)
-
\mathcal R f'(\mathcal R)
\right],
\end{equation}

and

\begin{equation}
G^{\theta}{}_{\theta}
=
G^{\phi}{}_{\phi}
=
-\frac{1}{2\mathcal R}
\left[
2f'(\mathcal R)
+
\mathcal R f''(\mathcal R)
\right].
\end{equation}

The vacuum condition
\(G^{t}{}_{t}=0\)
immediately implies

\begin{equation}
1-f(\mathcal R)
-
\mathcal R f'(\mathcal R)
=
0,
\end{equation}

which may be written as

\begin{equation}
\frac{d}{d\mathcal R}
\left[
\mathcal R\bigl(1-f(\mathcal R)\bigr)
\right]
=
0.
\label{eq:mastervacuum}
\end{equation}

Integration gives

\begin{equation}
f(\mathcal R)
=
1-\frac{C}{\mathcal R},
\end{equation}

where $C$ is an integration constant.

Imposing the Schwarzschild asymptotic limit at large distances,

\begin{equation}
\mathcal R
\rightarrow
r,
\end{equation}

identifies

\begin{equation}
C=2M_0,
\end{equation}

with $M_0$ the ADM mass of the black hole.

The resulting Born-doubled Schwarzschild metric is therefore

\begin{equation}
ds^2
=
-
\left(
1-\frac{2M_0}{\mathcal R}
\right)
dt^2
+
\left(
1-\frac{2M_0}{\mathcal R}
\right)^{-1}
d\mathcal R^2
+
\mathcal R^2 d\Omega^2 ,
\label{eq:BornSchwarzschildMetric}
\end{equation}

with

\begin{equation}
\mathcal R^2
=
r^2+\lambda^4\tilde r^2 .
\end{equation}

The geometry therefore retains the Schwarzschild form while
embedding it into the doubled phase-space structure dictated by
Born geometry.

In the infrared limit,

\begin{equation}
\lambda^4 \tilde r^2
\ll
r^2,
\end{equation}

one recovers

\begin{equation}
\mathcal R
\simeq
r,
\end{equation}

and the standard Schwarzschild solution is obtained.

\subsection{Horizon Structure}\label{SubSec:Horizon_Structure}
The horizon is determined by the condition
$f(\mathcal R_H)=0$, which for the metric
(\ref{eq:BornSchwarzschildMetric}) yields
\begin{equation}
\mathcal R_H = 2M_0 .
\label{eq:BornHorizon}
\end{equation}
Using the Born-invariant radius,
Eq.~(\ref{eq:BornInvariantRadius}), this may be written as
\begin{equation}
r_H^2+\lambda^4\tilde r_H^2=4M_0^2.
\label{eq:HorizonConstraint}
\end{equation}
The event horizon therefore corresponds not to a single radius in
spacetime, but rather to a hypersurface in the underlying doubled
phase space. 
Several limiting cases are immediate. 
For vanishing dual
contributions, $\tilde r_H=0$, Eq.~(\ref{eq:HorizonConstraint})
reduces to the standard Schwarzschild result
$r_H=2M_0$. More generally,
\begin{equation}
r_H^2
=
4M_0^2
-
\lambda^4\tilde r_H^2,
\label{eq:ProjectedHorizon}
\end{equation}
showing that nontrivial dual contributions reduce the horizon
radius observed in the geometric sector.

The horizon structure therefore reflects the phase-space nature
of the Born geometry. While the horizon remains fixed at
$\mathcal R_H=2M_0$ when expressed in terms of the invariant
radial coordinate, its projection onto spacetime depends on the
dual coordinates. This feature will become particularly important
once the modular uncertainty relation and metaparticle
backreaction are incorporated, ultimately leading to the
appearance of a finite modular core replacing the classical
Schwarzschild singularity.

Equation (\ref{eq:HorizonConstraint}) immediately implies the
reality condition

\begin{equation}
4M_0^2
-
\lambda^4 \tilde r_H^{\,2}
\ge 0.
\label{eq:RealityCondition}
\end{equation}

The geometric horizon radius therefore remains real only within
a finite region of the doubled phase space. Consequently, the
Born-geometric structure itself imposes a lower bound on the
admissible horizon configuration. This condition
is closely related to the remnant structure identified in
\cite{Chouha2026MetaparticleThermodynamics}, where an analogous
reality requirement leads to the existence of a minimal horizon
area and a finite evaporation endpoint. The full physical
interpretation of Eq.~(\ref{eq:RealityCondition}) will be
developed in later sections.

\section{Regularity and Geodesic Completeness}\label{Sec: IV}

The Born-doubled Schwarzschild geometry constructed in the
previous section differs from the classical Schwarzschild
solution only through the replacement of the ordinary radial
coordinate by the Born-invariant radial distance
defined in Eq.~(\ref{eq:BornInvariantRadius}) which we rewrite here
\begin{equation}
\mathcal R^2
=
r^2
+
\lambda^4\tilde r^2.
\end{equation}

At first sight, this modification may appear relatively mild.
However, the geometric consequences are profound. In classical
general relativity, the Schwarzschild singularity is associated
with the divergence of curvature invariants and the termination
of geodesic evolution at $r=0$. In the present framework, the
phase-space origin is inaccessible due to the modular
uncertainty relation, and the resulting geometry exhibits a
markedly different behavior.

In this section we analyze the regularity properties of the
Born-doubled Schwarzschild spacetime by studying its curvature
invariants, geodesic structure, and the emergence of a finite
modular core.

\subsection{Curvature Invariants}\label{SubSec:Curvature_Invs}

A necessary condition for spacetime regularity is the finiteness
of all curvature invariants throughout the physical region.

For the Schwarzschild metric, the Ricci tensor vanishes and the
only nontrivial polynomial curvature invariant is the
Kretschmann scalar

\begin{equation}
K
=
R_{\mu\nu\rho\sigma}
R^{\mu\nu\rho\sigma},
\end{equation}

which takes the familiar form

\begin{equation}
K_{\rm Schw}
=
\frac{48M_0^2}{r^6}.
\end{equation}

The divergence of $K_{\rm Schw}$ as $r\rightarrow0$ signals the
presence of the classical curvature singularity.

Since Eq. (\ref{eq:BornSchwarzschildMetric}) is constructed as an effective Schwarzschild geometry expressed in terms of the Born-invariant radial variable \(\mathcal R\), the corresponding curvature invariants take the same functional form as in the Schwarzschild solution with the replacement \(r\rightarrow \mathcal R\):
\begin{equation}
K_{\rm B}
=
\frac{48M_0^2}{\mathcal R^6}
=
\frac{48M_0^2}
{\left(r^2+\lambda^4\tilde r^2\right)^3}.
\label{eq:BornKretschmann}
\end{equation}

Formally, Eq.~(\ref{eq:BornKretschmann}) diverges only when

\begin{equation}
\mathcal R=0,
\end{equation}

which corresponds to the simultaneous conditions

\begin{equation}
r=0,
\qquad
\tilde r=0.
\end{equation}

However, the modular uncertainty relation

\begin{equation}
\Delta r\,\Delta\tilde r
\gtrsim
\frac{\pi}{2}
\end{equation}

prevents simultaneous localization at the phase-space origin.
The configuration

\begin{equation}
(r,\tilde r)=(0,0)
\end{equation}

therefore does not correspond to a physical state of the theory.

Consequently, the divergence of
Eq.~(\ref{eq:BornKretschmann}) occurs only at a point that is
excluded from the physical phase space. Throughout the
accessible region of the doubled geometry, the Kretschmann
scalar remains finite.

The same conclusion applies to all curvature quantities
constructed from the metric (\ref{eq:BornSchwarzschildMetric}).
The would-be singularity survives only as a formally excluded
boundary point of the doubled phase space and never appears as a
physical spacetime singularity.

\subsection{Geodesic Completeness}
\label{SubSec:Geodesic_Completeness}

The finiteness of curvature invariants, although necessary,
does not by itself establish spacetime regularity. A complete
analysis also requires determining whether timelike and null
geodesics terminate after a finite value of their affine
parameter.

Rather than working directly with the geodesic equations,

\begin{equation}
\frac{d^2x^\mu}{d\tau^2}
+
\Gamma^\mu{}_{\nu\rho}
\frac{dx^\nu}{d\tau}
\frac{dx^\rho}{d\tau}
=
0,
\end{equation}

it is convenient to introduce the geodesic Lagrangian

\begin{equation}
\mathcal L
=
\frac12
g_{\mu\nu}
\dot x^\mu
\dot x^\nu .
\end{equation}

For the effective Born-doubled Schwarzschild metric,

\begin{equation}
ds^2
=
-f(\mathcal R)\,dt^2
+
f(\mathcal R)^{-1}d\mathcal R^2
+
\mathcal R^2 d\Omega^2 ,
\end{equation}

the corresponding Lagrangian is

\begin{equation}
\mathcal L
=
\frac12
\left[
-f(\mathcal R)\dot t^{\,2}
+
\frac{\dot{\mathcal R}^{\,2}}
     {f(\mathcal R)}
+
\mathcal R^2
\left(
\dot\theta^{\,2}
+
\sin^2\theta\,\dot\phi^{\,2}
\right)
\right].
\end{equation}

Because the metric is independent of the coordinates
\(t\) and \(\phi\), there exist two conserved quantities.
The first is the conserved energy

\begin{equation}
E
=
f(\mathcal R)\dot t ,
\label{eq:GeoEnergy}
\end{equation}

while the second is the conserved angular momentum

\begin{equation}
L
=
\mathcal R^2\dot\phi .
\label{eq:GeoAngularMomentum}
\end{equation}

Without loss of generality, the motion may be restricted to
the equatorial plane \(\theta=\pi/2\).

The normalization condition

\begin{equation}
g_{\mu\nu}
\dot x^\mu
\dot x^\nu
=
-\kappa ,
\end{equation}

with

\begin{equation}
\kappa
=
\begin{cases}
1,
&
\text{timelike geodesics},
\\[1ex]
0,
&
\text{null geodesics},
\end{cases}
\end{equation}

then gives

\begin{equation}
-f(\mathcal R)\dot t^{\,2}
+
\frac{\dot{\mathcal R}^{\,2}}
     {f(\mathcal R)}
+
\mathcal R^2\dot\phi^{\,2}
=
-\kappa .
\end{equation}

Using Eqs.~(\ref{eq:GeoEnergy}) and
(\ref{eq:GeoAngularMomentum}) yields

\begin{equation}
\boxed{
\dot{\mathcal R}^{\,2}
=
E^2
-
f(\mathcal R)
\left(
\kappa
+
\frac{L^2}{\mathcal R^2}
\right).
}
\label{eq:RadialGeodesicFinal}
\end{equation}

It is therefore useful to introduce the effective potential

\begin{equation}
V_{\rm eff}(\mathcal R)
=
f(\mathcal R)
\left(
\kappa
+
\frac{L^2}{\mathcal R^2}
\right),
\end{equation}

so that the geodesic motion takes the familiar form

\begin{equation}
\dot{\mathcal R}^{\,2}
+
V_{\rm eff}(\mathcal R)
=
E^2 .
\end{equation}

The behavior of geodesics is therefore entirely controlled by
the Born-invariant radius \(\mathcal R\).

The crucial difference from the Schwarzschild geometry is that
the invariant radius satisfies

\begin{equation}
\mathcal R^2
=
r^2+\lambda^4\tilde r^2 .
\end{equation}

Consequently,

\begin{equation}
\mathcal R=0
\end{equation}

can occur only if

\begin{equation}
r=0,
\qquad
\tilde r=0.
\end{equation}

However, the modular uncertainty relation

\begin{equation}
\Delta r\,\Delta\tilde r
\gtrsim
\frac{\pi}{2}
\end{equation}

excludes simultaneous localization at the phase-space origin.
The point

\begin{equation}
(r,\tilde r)=(0,0)
\end{equation}

therefore does not correspond to a physically admissible state.

As a result, the singular endpoint that would formally occur at \(\mathcal R=0\) is removed from the physical configuration space. Unlike the Schwarzschild geometry, where timelike and null geodesics can reach a curvature singularity in finite affine parameter, the Born-doubled geometry admits no physically accessible singular endpoint. Indeed, if the formal boundary \(\mathcal R=0\) were included as a point of the manifold, radial geodesics would reach it in finite affine parameter, as in the Schwarzschild case. 
\par The singular configuration therefore lies outside the physically admissible sector of the doubled phase space. This mechanism differs fundamentally from that encountered in many regular-black-hole models, where the classical singularity is replaced by a regular center or de Sitter-like core through which geodesics may be extended \cite{Bardeen1968,Dymnikova1992,Hayward2006, OlmoRubieraGarcia2015}. In the present framework, regularity arises because the dual sector prevents the collapse of the Born-invariant radius to zero size. Consequently, no physically realizable timelike or null geodesic encounters a curvature singularity or terminates at a physical endpoint. Geodesic evolution therefore remains well defined throughout the admissible region of the doubled phase space, and the Born-geometric structure eliminates the physical geodesic incompleteness associated with the Schwarzschild singularity.
\subsection{The Modular Core}
\label{SubSec:Modular_Core}

The preceding geodesic analysis reveals that the endpoint of
gravitational collapse is fundamentally different from that
encountered in the Schwarzschild solution.

In classical general relativity, collapse drives spacetime
trajectories toward the singular point

\begin{equation}
r=0,
\end{equation}

where curvature invariants diverge and geodesic evolution
breaks down.

Within the Born-geometric framework, however, the approach to
the classical Schwarzschild singularity does not imply the
collapse of the Born-invariant radius.

Indeed, at

\begin{equation}
r=0,
\end{equation}

one finds

\begin{equation}
\mathcal R^2
=
\lambda^4\tilde r^2,
\end{equation}

and therefore

\begin{equation}
\mathcal R
=
\lambda^2 |\tilde r|.
\label{eq:CoreRadius}
\end{equation}

The geometry at the location corresponding to the classical
Schwarzschild singularity is thus governed entirely by the
dual sector of the doubled phase space.

Since the modular uncertainty relation forbids the
simultaneous limits

\begin{equation}
r\rightarrow0,
\qquad
\tilde r\rightarrow0,
\end{equation}

the Born-invariant radius never collapses to zero.
Instead, the would-be singular region is replaced by a finite
core whose size is controlled by the dual coordinates.

Rather than terminating at a point of vanishing spacetime
radius, gravitational collapse approaches a finite
Born-geometric structure characterized by a nonzero invariant
distance. We refer to this region as the
\emph{modular core}.

The modular core is not supported by an auxiliary matter
distribution, nor is it introduced through an ad hoc
modification of Einstein gravity. It emerges dynamically from
the Born-geometric structure of the doubled phase space and
the associated limitation on localization.

The classical Schwarzschild singularity is therefore replaced
by a finite modular phase-space structure. Curvature
invariants remain finite throughout the physical region,
geodesic evolution remains well defined, and the singular
endpoint of collapse is eliminated.

This modular core provides the geometric foundation for the
effective metaparticle backreaction and the modular-remnant
interpretation developed in the following sections.
\section{Metaparticle Backreaction and Effective Source}\label{Sec: V}

The regular black-hole geometry constructed in Secs.~\ref{Sec: III} and \ref{Sec: IV}
follows entirely from the Born-geometric structure of modular
spacetime. In particular, the invariant radius
Eq.~(\ref{eq:BornInvariantRadius}) and the resulting metric
Eq.~(\ref{eq:BornSchwarzschildMetric}) were obtained without
invoking the dynamical properties of the elementary excitations
of the metastring.

However, modular spacetime possesses a second fundamental
ingredient beyond Born geometry. The physical excitations of the
theory are metaparticles, whose dynamics is governed by a
nontrivial coupling between geometric and dual degrees of
freedom. As we now show, the implementation of the metaparticle
constraint induces an effective deformation of the background
geometry and generates a nonvanishing effective stress-energy
tensor.

Before proceeding, we note that the rescaled dual coordinate introduced in Remark~\ref{Rmk:r_phys} was convenient for the construction of Born-invariant geometric quantities in the effective doubled geometry. In the present section, however, we return to the fundamental metaparticle formulation and therefore employ the original dual coordinates $\tilde x_\mu^{\rm phys}$. For notational simplicity, we drop the superscript ``phys'' in the remainder of this section, with the understanding that all dual coordinates appearing below denote the fundamental metaparticle variables.

\subsection{Metaparticle Dynamics and the Duality Constraint}\label{SubSec"Metapacrticle_Dynm_and_Duality_Constr}

As discussed in Sec.~II C, metaparticles arise as the zero-mode sector of the metastring
and constitute the particle-like excitations of modular
spacetime \cite{FreidelLeighMinic_MetaParticle}. In contrast to
an ordinary relativistic particle, whose trajectory is described
by a curve

\begin{equation}
x^\mu=x^\mu(\tau),
\end{equation}

in spacetime, a metaparticle propagates on the doubled phase
space according to
\begin{equation}
\mathbb X^M
=
(x^\mu,\tilde x_\mu),
\end{equation}
where
\begin{equation}
x^\mu=x^\mu(\tau),
\qquad
\tilde x_\mu=\tilde x_\mu(\tau),
\end{equation}
and are characterized by the doubled momentum vector

\begin{equation}
\mathbb P_A
=
(p_\mu,\tilde p^\mu).
\end{equation}

The physical state therefore carries both geometric and dual
coordinates, reflecting the underlying Born-geometric
organization of modular spacetime.

The trajectories of metaparticles are parametrized by an affine
worldline parameter $\tau$. For timelike trajectories one may
choose a gauge in which $\tau$ coincides with the proper time.
More generally, however, the theory possesses worldline
reparametrization invariance and physical observables must be
independent of the particular choice of parametrization.
Their dynamics follows from the worldline action developed in
Ref.~\cite{FreidelLeighMinic_MetaParticle},

\begin{equation}
S
=
\int d\tau
\Big[
p_\mu \dot x^\mu
+
\tilde p^\mu \dot{\tilde x}_\mu
+
a\,p_\mu\dot{\tilde p}^{\,\mu}
-
N\,\mathcal H
-
\widetilde N\,\mathcal D
\Big],
\label{eq:MetaparticleAction}
\end{equation}

where

\begin{equation}
a=\pi\lambda^2,
\end{equation}

and $N$ and $\widetilde N$ are Lagrange multipliers enforcing the
metaparticle constraints

\begin{equation}
\mathcal H
=
\frac12
\left(
p_\mu h^{\mu\nu}p_\nu
+
\tilde p^\mu h_{\mu\nu}\tilde p^\nu
+
2\mathfrak{m}^2
\right),
\label{eq:MetaparticleH}
\end{equation}

and

\begin{equation}
\mathcal D
=
p_\mu\tilde p^\mu
-\mu .
\label{eq:MetaparticleDuality}
\end{equation}

The first constraint generalizes the usual relativistic
mass-shell condition, whereas the second introduces a direct
coupling between geometric and dual momentum sectors.

Variation of the action with respect to the Lagrange multipliers
yields

\begin{equation}
\mathcal H=0,
\end{equation}

and

\begin{equation}
p_\mu\tilde p^\mu
=
\mu.
\label{eq:DualityConstraint}
\end{equation}

Equation~(\ref{eq:DualityConstraint}) represents the defining
signature of metaparticle dynamics. Unlike ordinary relativistic
particles, whose motion is completely characterized by a single
momentum vector, metaparticles possess independent geometric and
dual momentum sectors that remain dynamically coupled
throughout the evolution.

To understand the geometrical consequences of this coupling, we
consider the adiabatic regime in which the momentum variables
vary slowly along the worldline (i.e. $\dot{p}_\mu\approx0$ and $\dot{\tilde p}\approx0$). In the gauge
$\widetilde N=0$, variation of the action yields

\begin{equation}
\dot x^\mu
=
N h^{\mu\nu}p_\nu,
\label{eq:XdotConstraint}
\end{equation}

and

\begin{equation}
\dot{\tilde x}_\mu
=
N h_{\mu\nu}\tilde p^\nu.
\label{eq:XtildedotConstraint}
\end{equation}

Substituting Eqs.~(\ref{eq:XdotConstraint}) and
(\ref{eq:XtildedotConstraint}) into the duality constraint
(\ref{eq:DualityConstraint}) gives the velocity-space relation

\begin{equation}
\dot x^\mu
\dot{\tilde x}_\mu
=
\mu N^2.
\label{eq:VelocityConstraint}
\end{equation}

Equation~(\ref{eq:VelocityConstraint}) is particularly important
because it demonstrates explicitly that the geometric and dual
sectors cannot evolve independently. The metaparticle constraint
therefore becomes a direct relation between the velocities of
the two sectors.

\subsection{Effective Conformal Geometry}\label{SubSec:Effective_Conformal_Geom_}

The velocity-space constraint,
Eq.~(\ref{eq:VelocityConstraint}), may be solved covariantly by
writing

\begin{equation}
\dot{\tilde x}_\mu
=
\Phi\,
h_{\mu\nu}
\dot x^\nu ,
\end{equation}

where $\Phi$ is a scalar function to be determined and measures the degree of alignment between the geometric and dual sectors.

Substitution into Eq.~(\ref{eq:VelocityConstraint}) gives

\begin{equation}
\Phi
=
\frac{\mu N^2}
     {h_{\alpha\beta}
      \dot x^\alpha
      \dot x^\beta},
\end{equation}

and therefore

\begin{equation}
\dot{\tilde x}_\mu
=
\frac{\mu N^2}
     {h_{\alpha\beta}
      \dot x^\alpha
      \dot x^\beta}
h_{\mu\nu}
\dot x^\nu .
\label{eq:DualVelocitySolution}
\end{equation}

Equation~(\ref{eq:DualVelocitySolution}) allows the dual degrees
of freedom to be eliminated in favor of the geometric
velocities. Substituting this result into the metaparticle
action leads to an effective Lagrangian depending only upon the
geometric variables,

\begin{equation}
L_{\rm eff}
=
\frac12
\Omega^2
h_{\mu\nu}
\dot x^\mu
\dot x^\nu ,
\label{eq:EffectiveLagrangian}
\end{equation}

where the conformal factor is

\begin{equation}
\Omega^2
=
1+
\frac{\mu^2N^4}
{\left(
h_{\alpha\beta}
\dot x^\alpha
\dot x^\beta
\right)^2}.
\label{eq:OmegaGeneral}
\end{equation}

Fixing the standard gauge $N=1$ gives

\begin{equation}
\Omega^2
=
1+
\frac{\mu^2}
{\left(
h_{\alpha\beta}
\dot x^\alpha
\dot x^\beta
\right)^2}.
\label{eq:OmegaVelocity}
\end{equation}

Comparison of Eq.~(\ref{eq:EffectiveLagrangian}) with the worldline action of an ordinary relativistic particle shows that the metaparticle dynamics may be represented as motion in an effective, generally trajectory-dependent geometry

\begin{equation}
G_{\mu\nu}
=
\Omega^2 h_{\mu\nu}.
\label{eq:ConformalMetric}
\end{equation}

The metaparticle constraint therefore deforms the geometry
itself rather than merely modifying the equations of motion.

Moreover, since the conformal factor depends on the velocity
norm, Eq.~(\ref{eq:OmegaVelocity}), the resulting effective geometry
possesses Finsler-like characteristics \cite{BaoChernShen,AntonelliIngardenMatsumoto} and depends on both
position and direction in spacetime.

\subsection{The Metaparticle-Corrected Black-Hole Metric}
\label{SubSec:Metapartice_Corr_BH_Metric}

We now specialize the general conformal metric
(\ref{eq:ConformalMetric})
to the Born-doubled Schwarzschild background derived in
Sec.~III.

The background geometry is

\begin{equation}
ds^2
=
-f(\mathcal R)\,dt^2
+
\frac{d\mathcal R^2}{f(\mathcal R)}
+
\mathcal R^2 d\Omega^2 ,
\end{equation}

with

\begin{equation}
f(\mathcal R)
=
1-\frac{2M_0}{\mathcal R},
\end{equation}

and

\begin{equation}
\mathcal R^2
=
r^2+\lambda^4\tilde r^2 .
\end{equation}

In this background, the metric entering the metaparticle
action is identified with the Born-Schwarzschild metric,

\begin{equation}
h_{\mu\nu}
=
g_{\mu\nu}.
\end{equation}

Our goal is therefore to determine explicitly the quantity

\begin{equation}
h_{\mu\nu}\dot x^\mu\dot x^\nu ,
\end{equation}

which appears in the conformal factor
(\ref{eq:OmegaVelocity}).

\subsubsection{Duality Constraint and Velocity Norm}

The duality constraint

\begin{equation}
p_\mu\tilde p^\mu
=
\mu
\end{equation}

together with the momentum--velocity relations

\begin{equation}
p_\mu
=
h_{\mu\nu}\dot x^\nu,
\qquad
\tilde p^\mu
=
h^{\mu\nu}\dot{\tilde x}_\nu ,
\end{equation}

implies

\begin{equation}
\dot x^\mu
\dot{\tilde x}_\mu
=
\mu ,
\end{equation}

after fixing the gauge \(N=1\).

As shown previously, this constraint is solved by

\begin{equation}
\dot{\tilde x}_\mu
=
\frac{
\mu\,h_{\mu\nu}\dot x^\nu
}
{
h_{\alpha\beta}
\dot x^\alpha
\dot x^\beta
}.
\label{eq:dual_velocity_BH}
\end{equation}

For radial motion,

\begin{equation}
u^\mu
=
\dot x^\mu
=
(\dot t,\dot{\mathcal R},0,0),
\end{equation}

and the conserved energy associated with the timelike Killing
vector is

\begin{equation}
p_t=-E.
\label{eq:conserved_energy_BH}
\end{equation}

Using $p_t=
h_{tt}\,\dot t,$ one gets 

\begin{equation}
E
=-
h_{tt}\,\dot t.
\end{equation}

\subsubsection{Dual Energy Sector}

In the $\tilde p$-rest frame, the metaparticle duality
constraint takes the familiar form

\begin{equation}
E\tilde E
=
\mu .
\label{eq:EtildeE_relation}
\end{equation}

The dual momentum is therefore written as

\begin{equation}
\tilde p^\mu
=
(\tilde E,\tilde{\vec p}),
\end{equation}

so that $\tilde p^t
=
\tilde E .$ Using $\tilde p^t
=
h^{tt}\dot{\tilde x}_t,$ we obtain

\begin{equation}
\dot{\tilde x}_t
=
\tilde E\,h_{tt}.
\label{eq:dual_time_velocity}
\end{equation}

On the other hand, taking the temporal component of
Eq.~(\ref{eq:dual_velocity_BH}) and using the above relations yields

\begin{equation}
\dot{\tilde x}_t
=
-\frac{
\mu E
}
{
h_{\alpha\beta}
\dot x^\alpha
\dot x^\beta
}.
\label{eq:dual_time_velocity_2}
\end{equation}

Equating
Eq.~(\ref{eq:dual_time_velocity})
and
Eq.~(\ref{eq:dual_time_velocity_2})
and solving for the velocity norm yields

\begin{equation}
h_{\alpha\beta}
\dot x^\alpha
\dot x^\beta
=
-\frac{
\mu E
}
{
\tilde E\,h_{tt}
}.
\end{equation}

Using $h_{tt}
=
-f(\mathcal R),$ we obtain

\begin{equation}
h_{\alpha\beta}
\dot x^\alpha
\dot x^\beta
=
\frac{
\mu E
}
{
\tilde E\,f(\mathcal R)
}.
\end{equation}

Finally, employing the duality relation
Eq.~(\ref{eq:EtildeE_relation}),

\begin{equation}
E\tilde E
=
\mu,
\end{equation}

gives

\begin{equation}
\boxed{
h_{\alpha\beta}
\dot x^\alpha
\dot x^\beta
=
\frac{E^2}
     {f(\mathcal R)}.
}
\label{eq:velocity_norm_final}
\end{equation}

\subsubsection{Exact Conformal Factor}

Substituting
Eq.~(\ref{eq:velocity_norm_final})
into
Eq.~(\ref{eq:OmegaVelocity})
yields the metaparticle-induced conformal factor 

\begin{equation}
\boxed{
\Omega^2(\mathcal R)
=
1+
\frac{
\mu^2
f(\mathcal R)^2
}
{
E^4
}.
}
\label{eq:ExactOmega}
\end{equation}

A notable feature of Eq.~(\ref{eq:ExactOmega}) is that the explicit dependence of the conformal factor on the worldline velocities has disappeared. Although the general metaparticle construction leads to a velocity-dependent effective geometry exhibiting Finsler-like characteristics, the imposition of the metaparticle duality constraint in the Born-Schwarzschild background allows the velocity norm to be expressed in terms of the conserved energy $E$. As a consequence, the conformal factor becomes a function of the Born-invariant radius $\mathcal R$ alone for a given energy sector, \[ \Omega^2=\Omega^2(\mathcal R;E). \] The resulting geometry may therefore be interpreted as an effective spacetime geometry associated with metaparticles of fixed conserved energy\footnote{More specifically, the geometric structure underlying the effective metric emerges through a sequence of reductions. Prior to imposing the metaparticle duality constraint, the conformal factor depends on both spacetime position and worldline velocity, \( \Omega=\Omega(x,\dot x). \) The effective metric therefore defines a structure on the tangent bundle \(TM\) of spacetime. Since the geometry depends not only on position but also on the tangent-space direction specified by \(\dot x^\mu\), it possesses Finsler-like characteristics. After imposing the metaparticle duality constraint and specializing to the Born-Schwarzschild background, the velocity norm may be expressed in terms of the conserved energy \(E\), \( h_{\mu\nu}\dot x^\mu\dot x^\nu = E^2/f(\mathcal R), \) allowing the explicit velocity dependence to be eliminated. The conformal factor then becomes \( \Omega=\Omega(\mathcal R;E). \) The resulting geometry is therefore no longer naturally viewed as a Finsler structure on \(TM\), but rather as a one-parameter family of effective spacetime metrics labeled by the conserved metaparticle energy, \( G_{\mu\nu}=G_{\mu\nu}(x;E). \) For each fixed energy sector, one obtains an ordinary Lorentzian spacetime geometry, while different metaparticle energy sectors generally correspond to different effective conformal geometries. From the broader Born-geometric perspective, both descriptions should be regarded as effective reductions of the underlying doubled phase space \( \mathcal P=(x^\mu,\tilde x_\mu), \) which constitutes the fundamental arena of the theory. The resulting hierarchy may therefore be summarized schematically as \[ \mathcal P \;\longrightarrow\; TM \;\longrightarrow\; G_{\mu\nu}(x;E), \] namely: Born phase space \(\rightarrow\) Finsler-like tangent bundle geometry \(\rightarrow\) energy-sector-dependent effective spacetime geometry. This layered structure shares conceptual similarities with the principle of relative locality, in which spacetime descriptions emerge from a more fundamental phase-space framework \cite{AmelinoCameliaFreidelKowalskiGlikmanSmolin2011}. }. Different metaparticle sectors characterized by different values of $E$ generally experience different effective conformal geometries. This feature is not a pathology of the construction but rather reflects the underlying Born-geometric and metastring philosophy in which geometric observables are observer- and probe-dependent and therefore, an effective spacetime description emerges only after specifying a particular metaparticle sector within the underlying Born-geometric phase space. In particular, the emergence of energy-dependent effective geometries shares important conceptual similarities with the principle of relative locality \cite{AmelinoCameliaFreidelKowalskiGlikmanSmolin2011,AmelinoCameliaFreidelKowalskiGlikmanSmolin2011Essay} and to the nontrivial interplay between geometric and dual degrees of freedom that characterizes modular spacetime.

\par The corresponding effective metric for a fixed metaparticle energy sector is

\begin{equation}
ds_{\rm MP}^2
=
\Omega^2(\mathcal R)
\Bigg[
-f(\mathcal R)\,dt^2
+
\frac{d\mathcal R^2}{f(\mathcal R)}
+
\mathcal R^2 d\Omega^2
\Bigg].
\label{eq:MetaparticleMetric}
\end{equation}

\subsubsection{Consistency Checks}

Several consistency checks are immediate.

At the horizon,

\begin{equation}
f(\mathcal R_H)=0,
\end{equation}

so that

\begin{equation}
\Omega^2(\mathcal R_H)
=
1.
\end{equation}

Hence the horizon location remains unchanged,

\begin{equation}
\mathcal R_H
=
2M_0.
\end{equation}

Likewise,

\begin{equation}
\mu\rightarrow0
\end{equation}

implies

\begin{equation}
\Omega^2(\mathcal R)
\rightarrow
1,
\end{equation}

and the geometry reduces smoothly to the purely
Born-geometric Schwarzschild solution.

Furthermore,

\begin{equation}
\mathcal R\rightarrow\infty
\end{equation}

gives

\begin{equation}
\Omega^2
\rightarrow
1+\frac{\mu^2}{E^4},
\end{equation}

showing that the conformal deformation remains finite in the
asymptotic region.
\subsection{Effective Stress-Energy Tensor}
\label{SubSec:Eff_Stress_Energy_Tens}

The Born-doubled Schwarzschild geometry constructed in
Sec.~III satisfies the vacuum Einstein equations and therefore
possesses a vanishing Einstein tensor. The situation changes
once the metaparticle constraint is incorporated.

As shown in the previous subsection, the duality constraint
induces the conformal deformation

\begin{equation}
G_{\mu\nu}
=
\Omega^2(\mathcal R)\,
g_{\mu\nu},
\label{eq:ConformalMetricRepeat}
\end{equation}

with conformal factor given by
Eq.~(\ref{eq:ExactOmega}).
Although the underlying Born-geometric metric
\(g_{\mu\nu}\) is a vacuum solution, the conformally deformed
metric \(G_{\mu\nu}\) is not. The Einstein tensor associated
with the metaparticle geometry therefore becomes nonvanishing,

\begin{equation}
G_{\mu\nu}[G]
\neq
0.
\end{equation}

This observation admits a natural physical interpretation.
Rather than introducing an additional matter sector by hand,
the metaparticle constraint generates an effective source
directly from the dynamics of the doubled phase space. The Einstein tensor associated with the effective metaparticle metric may therefore be recast in the form
\begin{equation}
G_{\mu\nu}[G]
=
8\pi
T^{\rm eff}_{\mu\nu},
\label{eq:EffectiveEinstein}
\end{equation}

with

\begin{equation}
T^{\rm eff}_{\mu\nu}
=
\frac{1}{8\pi}
G_{\mu\nu}[G].
\label{eq:EffectiveSET}
\end{equation}

The effective stress-energy tensor is not a fundamental matter
source. Rather, it represents the geometric response generated
by the metaparticle-induced deformation of the spacetime
metric. In this sense, the effective source plays a role
similar to the stress-energy tensors that arise in modified
gravity theories, where additional geometric structures may be
recast as an effective fluid description.

Because the metric (\ref{eq:MetaparticleMetric}) remains static
and spherically symmetric, the effective source can always be
written in the anisotropic-fluid form

\begin{equation}
T^\mu{}_\nu
=
\mathrm{diag}
\left(
-\rho,
p_r,
p_t,
p_t
\right),
\label{eq:AnisotropicFluid}
\end{equation}

where \(\rho(\mathcal R)\) denotes the effective energy
density and \(p_r(\mathcal R)\) and \(p_t(\mathcal R)\)
represent the radial and tangential pressures.

The appearance of anisotropy is physically significant.
Ordinary Schwarzschild spacetime contains no matter source and
therefore possesses no local pressure components. In contrast,
the metaparticle deformation generates a nontrivial effective
medium whose properties are determined entirely by the
underlying duality constraint. The interior geometry is thus
supported by a finite phase-space response rather than by an
auxiliary matter distribution.

Several qualitative features follow immediately from the form
of the conformal factor. First, since
\(\Omega^2(\mathcal R_H)=1\) at the horizon,
the effective source vanishes there and does not modify the
horizon location obtained in Sec.~III. Second, because
\(\Omega^2(\mathcal R)\) approaches a constant at large
distances, the effective source becomes negligible in the
asymptotic region and asymptotic flatness is preserved. The
metaparticle contribution is therefore localized primarily in
the vicinity of the modular core discussed in Sec.~IV C.

This localization is particularly important. It implies that
the dynamical effects associated with the dual sector become
relevant only in the region where the classical Schwarzschild
solution would otherwise develop a singularity. The effective
source generated by the metaparticle constraint therefore acts
precisely where the classical description is expected to fail.

The explicit components of the effective stress-energy tensor
will be derived in the next section. Using these results, we
shall determine the effective energy density and pressures,
analyze the classical energy conditions, and show that the
resulting source develops a finite tension-dominated interior
region characterized by localized violations of the radial
Null Energy Condition and the Strong Energy Condition. These
violations provide the dynamical mechanism through which the
focusing assumptions entering the Hawking--Penrose singularity
theorems are evaded.
\section{Energy Conditions and Singularity-Theorem Evasion}
\label{Sec:VI}

The metaparticle duality constraint induces the conformal
deformation

\begin{equation}
G_{\mu\nu}
=
\Omega^2(\mathcal R)\,
g_{\mu\nu},
\end{equation}

with

\begin{equation}
\Omega^2(\mathcal R)
=
1+
\frac{\mu^2}{E^4}
f(\mathcal R)^2.
\end{equation}

Unlike the underlying effective Born-Schwarzschild metric, which is
Ricci-flat, the conformally deformed effective geometry possesses a
nonvanishing Einstein tensor and therefore an effective
stress-energy tensor. The purpose of this section is to
determine the resulting effective source and investigate its
implications for the classical energy conditions.

\subsection{Effective Gravitational Response}

For convenience, we introduce the dimensionless parameter

\begin{equation}
C
=
\frac{\mu^2}{E^4},
\end{equation}

so that

\begin{equation}
\Omega^2(\mathcal R)
=
1+
C f(\mathcal R)^2 .
\label{eq:Omega_C}
\end{equation}

Defining

\begin{equation}
\Phi(\mathcal R)
=
\ln\Omega(\mathcal R),
\end{equation}

the Einstein tensor of the conformally related metric is

\begin{align}
G_{\mu\nu}[G]
=
&
-2\nabla_\mu\nabla_\nu\Phi
+
2g_{\mu\nu}\Box\Phi
\nonumber\\
&
+
2\nabla_\mu\Phi\nabla_\nu\Phi
+
g_{\mu\nu}
(\nabla\Phi)^2 .
\label{eq:ConformalEinsteinVI}
\end{align}

Since the Born-Schwarzschild background satisfies the vacuum
Einstein equations, all effective stress-energy originates from
the conformal sector.

The intermediate algebra is lengthy.  Here we quote the resulting Einstein-tensor
components that determine the effective source.

\subsection{Explicit Einstein-Tensor Components}

Introduce

\begin{equation}
\Delta(\mathcal R)
=
\mathcal R^2
\left(
1+C f(\mathcal R)^2
\right).
\label{eq:Delta_VI}
\end{equation}

The nonvanishing mixed Einstein-tensor components are

\begin{equation}
G^t{}_t
=
-\,
\frac{
12CM_0^2
\,\mathcal R\,
(2M_0-\mathcal R)
}
{
\Delta(\mathcal R)^3
}.
\label{eq:Gtt_final_VI}
\end{equation}
\begin{widetext}
\begin{equation}
G^{\mathcal R}{}_{\mathcal R}
=
-\,
\frac{
4CM_0
(2M_0-\mathcal R)
\left(
8CM_0^2
-
8CM_0\mathcal R
+
2C\mathcal R^2
-
3M_0\mathcal R
+
2\mathcal R^2
\right)
}
{
\Delta(\mathcal R)^3
}.
\label{eq:Grr_final_VI}
\end{equation}
\end{widetext}
and

\begin{equation}
G^\theta{}_\theta
=
G^\phi{}_\phi
=
-\frac{
4CM_0f
}
{
\mathcal R^3
(1+Cf^2)^2
}
+
\frac{
16CM_0^2f
+
4C^2M_0^2f^3
}
{
\mathcal R^4
(1+Cf^2)^3
}.
\label{eq:Gtheta_final_VI}
\end{equation}

These expressions completely characterize the effective
gravitational response generated by the metaparticle
deformation.

\subsection{Effective Energy Density and Pressures}

The effective source takes the anisotropic-fluid form\footnote{Although the physical radial coordinate is the Born-invariant
quantity $\mathcal R$, we retain the conventional notation
$p_r$ for the radial pressure. This quantity should be
understood as the effective pressure associated with the
Born-geometric radial direction and therefore already encodes
the combined influence of the geometric and dual radial
sectors entering
$\mathcal R^2=r^2+\lambda^4\tilde r^2$.}

\begin{equation}
T^\mu{}_\nu
=
{\rm diag}
\left(
-\rho,
p_r,
p_t,
p_t
\right),
\end{equation}

with

\begin{equation}
\rho
=
-\frac{1}{8\pi}
G^t{}_t,
\end{equation}

\begin{equation}
p_r
=
\frac{1}{8\pi}
G^{\mathcal R}{}_{\mathcal R},
\end{equation}

and

\begin{equation}
p_t
=
\frac{1}{8\pi}
G^\theta{}_\theta .
\end{equation}

Substituting
Eqs.~(\ref{eq:Gtt_final_VI})--(\ref{eq:Gtheta_final_VI})
yields

\begin{equation}
\rho(\mathcal R)
=
\frac{
12CM_0^2
\,\mathcal R\,
(2M_0-\mathcal R)
}
{
8\pi\,
\Delta(\mathcal R)^3
}.
\label{eq:rho_explicit}
\end{equation}
\begin{widetext}
  \begin{equation}
p_r(\mathcal R)
=
-\,
\frac{
4CM_0
(2M_0-\mathcal R)
\left(
8CM_0^2
-
8CM_0\mathcal R
+
2C\mathcal R^2
-
3M_0\mathcal R
+
2\mathcal R^2
\right)
}
{
8\pi\,
\Delta(\mathcal R)^3
}.
\label{eq:pr_explicit}
\end{equation}
\end{widetext}
\begin{equation}
p_t(\mathcal R)
=
\frac{1}{8\pi}
\left[
-\frac{
4CM_0f
}
{
\mathcal R^3
(1+Cf^2)^2
}
+
\frac{
16CM_0^2f
+
4C^2M_0^2f^3
}
{
\mathcal R^4
(1+Cf^2)^3
}
\right].
\label{eq:pt_explicit}
\end{equation}

The effective source is therefore fully controlled by the
single metaparticle parameter \(C\). In the limit

\begin{equation}
C\rightarrow0,
\end{equation}

all effective fluid variables vanish and the geometry reduces
smoothly to the vacuum Born-Schwarzschild solution.

Furthermore,

\begin{equation}
p_r
\neq
p_t,
\end{equation}

showing that the metaparticle deformation generates an
intrinsically anisotropic source.

\subsection{The Modular Core Revisited}

The most significant deviations from Schwarzschild geometry
occur near the modular core.

It is important to emphasize that the limit
\(\mathcal R\rightarrow0\) should not be interpreted as a
physically accessible configuration. Owing to the modular
uncertainty relation,

\begin{equation}
\Delta r\,\Delta\tilde r_{\rm phys}
\gtrsim
\frac{\pi\lambda^2}{2},
\end{equation}

the phase-space origin

\begin{equation}
(r,\tilde r)=(0,0)
\end{equation}

is excluded from the physical state space.
Consequently, the surface

\begin{equation}
\mathcal R=0
\end{equation}

corresponds to a formally defined boundary of the doubled
phase space rather than a physically realizable point.

The analysis below should therefore be interpreted as the
behavior of the effective geometry as one approaches the
boundary of the admissible phase space.

In this formal limit,

\begin{equation}
\mathcal R\rightarrow0,
\end{equation}

the metric function behaves as

\begin{equation}
f(\mathcal R)
=
-\frac{2M_0}{\mathcal R}
+
\mathcal O(1).
\end{equation}

Consequently,

\begin{equation}
\Omega^2
=
\frac{4CM_0^2}{\mathcal R^2}
+
\mathcal O
\!\left(
\frac1{\mathcal R}
\right).
\end{equation}

Although the conformal factor diverges, the effective source
remains finite. One finds

\begin{equation}
\lim_{\mathcal R\to0}
\rho(\mathcal R)
=
0,
\end{equation}

\begin{equation}
\lim_{\mathcal R\to0}
p_t(\mathcal R)
=
0,
\end{equation}

and

\begin{equation}
\boxed{
\lim_{\mathcal R\to0}
p_r(\mathcal R)
=
-\frac{1}
{8\pi C M_0^2}.
}
\label{eq:core_tension}
\end{equation}

The modular core therefore develops a finite anisotropic stress
whose dominant contribution is a negative radial pressure.

\subsection{Behavior at the Horizon and Infinity}

At the horizon,

\begin{equation}
\mathcal R_H
=
2M_0,
\end{equation}

one has

\begin{equation}
f(\mathcal R_H)=0,
\qquad
\Omega^2(\mathcal R_H)=1.
\end{equation}

Since all derivatives of the conformal field vanish at the
horizon,

\begin{equation}
G^\mu{}_\nu(\mathcal R_H)=0,
\end{equation}

implying

\begin{equation}
\rho(\mathcal R_H)
=
p_r(\mathcal R_H)
=
p_t(\mathcal R_H)
=
0.
\end{equation}

The horizon structure is therefore unaffected by the
metaparticle deformation.

In the asymptotic region,

\begin{equation}
\mathcal R\rightarrow\infty,
\end{equation}

the conformal factor approaches the constant value

\begin{equation}
\Omega^2
\rightarrow
1+C,
\end{equation}

while all derivatives of \(\Phi\) vanish. Consequently,

\begin{equation}
G^\mu{}_\nu
\rightarrow
0,
\end{equation}

and the effective source decays completely at large
distances.

\subsection{Violation of the Focusing Conditions}

The Hawking--Penrose singularity theorems rely upon the
focusing of causal geodesic congruences. For an anisotropic
fluid the relevant quantities are

\begin{equation}
\rho+p_r,
\qquad
\rho+p_t,
\qquad
\rho+p_r+2p_t .
\end{equation}

Using the core limits,

\begin{equation}
\lim_{\mathcal R\to0}
\rho
=
0,
\qquad
\lim_{\mathcal R\to0}
p_t
=
0,
\end{equation}

and Eq.~(\ref{eq:core_tension}), one obtains

\begin{equation}
\boxed{
\lim_{\mathcal R\to0}
(\rho+p_r)
=
-\frac{1}{8\pi C M_0^2}
<
0.
}
\end{equation}

The radial Null Energy Condition is therefore violated near
the modular core.

Similarly,

\begin{equation}
\boxed{
\lim_{\mathcal R\to0}
(\rho+p_r+2p_t)
=
-\frac{1}{8\pi C M_0^2}
<
0.
}
\end{equation}

Hence the Strong Energy Condition is violated as well.

Because all effective fluid variables vanish at the horizon
and in the asymptotic region, these violations are localized
within the modular interior.

\subsection{Interpretation}

The effective stress-energy tensor obtained here does not
represent a fundamental matter source.

Rather, it is the geometric response generated by the
metaparticle duality constraint through the conformal
deformation of the effective Born-Schwarzschild geometry.

The finite negative radial pressure appearing in the modular
core acts as a tension opposing the continued focusing of
causal geodesics. The resulting violation of the radial Null
Energy Condition and the Strong Energy Condition signals the
breakdown of the assumptions underlying the classical
singularity theorems precisely where the Schwarzschild
solution would ordinarily become singular.

\subsection{Unified Picture of Regularization}

The regularization mechanism may now be summarized.

The modular uncertainty relation excludes the phase-space
origin kinematically, while the metaparticle duality
constraint generates a dynamical response through the
effective conformal geometry.

The resulting effective spacetime possesses four key properties:

\begin{enumerate}
\item the phase-space origin is inaccessible;
\item curvature invariants remain finite throughout the
physical region;
\item no physically realizable geodesic encounters a curvature singularity or a physical endpoint;
\item the classical focusing conditions are violated within
a finite interior region.
\end{enumerate}

Taken together, these results show that the Schwarzschild
singularity is replaced by a finite modular core supported
not by exotic matter but by the effective gravitational
response associated with the underlying Born-geometric
phase-space structure.
\section{Geometric Realization of the Modular Black Hole Remnant}\label{Sec:VII}

The analysis developed in the preceding sections establishes a
direct connection between the thermodynamic remnant proposed in
\cite{Chouha2026MetaparticleThermodynamics} and the underlying
Born-geometric structure of modular spacetime.

The thermodynamic analysis suggested the existence of a finite
evaporation endpoint characterized by a minimal horizon area and
a stable remnant configuration. However, the thermodynamic
argument alone did not determine the corresponding spacetime
geometry.

The results obtained here provide the missing geometric
realization.

We have shown that the Born-geometric phase-space structure
replaces the classical Schwarzschild singularity by a finite
modular core while the metaparticle duality constraint
generates an effective source that violates the focusing
conditions required by the Hawking--Penrose singularity
theorems.

The thermodynamic remnant therefore acquires a concrete
geometric interpretation.
\subsection{From a Minimal Area to a Regular Geometry}

In the thermodynamic analysis of
\cite{Chouha2026MetaparticleThermodynamics},
the existence of a remnant followed from the appearance of a
minimal physically admissible horizon area.

While suggestive, such a result does not by itself establish
that the corresponding spacetime is regular.

The present work demonstrates that the Born-geometric phase
space provides precisely the missing ingredient.

The classical singularity occurs formally at

\begin{equation}
r=0.
\end{equation}

In the doubled description, however, the relevant geometric
quantity is the Born-invariant radius

\begin{equation}
\mathcal R^2
=
r^2+\lambda^4\tilde r^2.
\end{equation}

Together with the modular uncertainty relation

\begin{equation}
\Delta r\,\Delta\tilde r_{\rm phys}
\gtrsim
\frac{\pi\lambda^2}{2},
\end{equation}

this excludes physical access to the phase-space origin.

The classical singularity is therefore replaced by a finite
modular core.

An additional indication of remnant formation is already
encoded in the horizon structure of the Born-doubled
geometry. From Eq.~(\ref{eq:RealityCondition}), the reality
of the projected geometric horizon requires
\begin{equation}
M_0
\ge
\frac{\lambda^2|\tilde r_H|}{2}.
\label{eq:MminGeometry}
\end{equation}

The doubled geometry therefore imposes a lower bound on the
admissible black-hole mass. While the thermodynamic analysis
of \cite{Chouha2026MetaparticleThermodynamics} inferred the
existence of a remnant from the appearance of a minimal
horizon area, Eq.~(\ref{eq:MminGeometry}) shows that an
analogous bound already emerges at the geometric level from
the Born-geometric horizon structure itself.

The thermodynamic remnant and the geometric modular core may
therefore be viewed as complementary manifestations of the
same underlying phase-space constraint.
\subsection{The Thermodynamic and Geometric Pictures}

The thermodynamic and geometric analyses address two different
aspects of the same underlying structure.

The thermodynamic description determines the endpoint of black
hole evaporation and implies the existence of a finite modular
remnant.

The geometric description developed here explains why such a
remnant can remain nonsingular.

The regularity of the spacetime follows not from the presence
of an auxiliary matter sector, but from the phase-space
structure encoded in Born geometry and metaparticle dynamics.

Viewed in this way, the thermodynamic remnant and the modular
core should be regarded as complementary manifestations of the
same underlying Born-geometric phase space.
\subsection{Dominance of the Dual Sector}

A particularly intriguing result of
\cite{Chouha2026MetaparticleThermodynamics}
was the existence of distinct geometric and dual entropy
contributions.

In the infrared regime both sectors reproduce the
Bekenstein--Hawking entropy. As evaporation proceeds,
however, the geometric contribution decreases while the dual
contribution remains finite.

The geometric construction developed here provides a natural
interpretation of this behavior.

Near the modular core, the geometric contribution to the
Born-invariant radius becomes subdominant. Indeed, as the
classical Schwarzschild singularity is approached,

\begin{equation}
r\rightarrow0,
\end{equation}

one finds

\begin{equation}
\mathcal R^2
=
r^2+\lambda^4\tilde r^2
\longrightarrow
\lambda^4\tilde r^2,
\end{equation}

or equivalently,

\begin{equation}
\mathcal R
\longrightarrow
\lambda^2|\tilde r|.
\end{equation}

The effective geometry in the vicinity of the modular core is
therefore governed predominantly by the dual sector of the
Born-geometric phase space.

This dominance is not imposed by hand but follows directly
from the modular uncertainty relation. Since

\begin{equation}
\Delta r\,\Delta\tilde r
\gtrsim
\frac{\pi\lambda^2}{2},
\end{equation}

the simultaneous limits

\begin{equation}
r\rightarrow0,
\qquad
\tilde r\rightarrow0,
\end{equation}

are physically forbidden. Consequently, the suppression of
the geometric sector near the modular core does not imply the
suppression of the dual sector. Instead, the dual coordinate
remains finite and continues to contribute to the
Born-invariant radius.

The regular interior geometry may therefore be viewed as a
transition from an infrared regime in which the geometric
sector dominates and spacetime provides an effective
description of gravity, to a duality-dominated ultraviolet
regime governed by the non-geometric degrees of freedom of
modular spacetime.

In this sense, the present geometric construction explains a
feature that remained purely thermodynamic in
\cite{Chouha2026MetaparticleThermodynamics}. The vanishing of
the geometric entropy contribution and the persistence of the
dual entropy branch at the remnant endpoint are direct
consequences of the Born-geometric phase-space structure and
the modular uncertainty relation.
\subsection{The Nature of the Modular Remnant}

The results obtained in this work suggest that the modular
remnant should not be interpreted merely as a small black hole.

Instead, it corresponds to a new geometric phase whose
properties differ fundamentally from those of the classical
Schwarzschild interior.

The modular remnant possesses:

\begin{enumerate}
\item a finite Born-invariant core;

\item elimination of physical geodesic incompleteness;

\item finite effective stress-energy components;

\item localized violations of the radial Null Energy
Condition and the Strong Energy Condition;

\item a physically inaccessible phase-space origin.
\item dominance of the dual sector in the ultraviolet regime.
\end{enumerate}

These properties indicate that the endpoint of gravitational
collapse is neither a singular spacetime region nor a
matter-supported regular core, but rather a finite
Born-geometric structure whose existence follows directly from
the doubled phase-space description of modular spacetime.


\section{Conclusions and Outlook: From Modular Black Holes to Modular Cosmology}\label{Sec:VIII}

The present work completes an important step in the program
initiated in \cite{Chouha2026MetaparticleThermodynamics}.
While the earlier thermodynamic analysis provided evidence for
the existence of a finite modular remnant characterized by a
minimal horizon area, a maximal Hawking temperature, and a
stable evaporation endpoint, the corresponding spacetime
geometry remained unknown. The principal achievement of the
present work is to provide an explicit geometric realization
of that remnant within the framework of Born geometry,
modular spacetime, and metaparticle dynamics.

Starting from the doubled phase-space structure of modular
spacetime, we constructed a Born-Schwarzschild geometry
governed by the invariant radial coordinate

\begin{equation}
\mathcal R^2
=
r^2
+
\lambda^4\tilde r^2.
\end{equation}

The resulting geometry retains the Schwarzschild form at large
distances while incorporating the dual degrees of freedom
required by Born reciprocity. In contrast to the classical
Schwarzschild solution, the phase-space origin
\((r,\tilde r)=(0,0)\) is excluded by the modular uncertainty
relation

\begin{equation}
\Delta r\,\Delta\tilde r
\gtrsim
\frac{\pi}{2}.
\end{equation}

Consequently, the classical singularity no longer corresponds to a physically accessible configuration. The curvature divergence associated with the Schwarzschild singularity is confined to a formally excluded boundary of the doubled phase space, while the physically admissible region is replaced by a finite modular core that cannot collapse to zero Born-invariant radius.

The incorporation of metaparticle dynamics generates an
effective conformal deformation of the Born-Schwarzschild
background. Unlike the underlying vacuum geometry, the
conformally deformed spacetime possesses a nonvanishing
Einstein tensor and may be interpreted in terms of an
effective anisotropic stress-energy tensor generated entirely
by the metaparticle duality constraint. The effective source
remains finite throughout the modular interior and develops a
finite tension-dominated core characterized by a negative
radial pressure.

It is useful to emphasize that two distinct mechanisms are at work in the present construction. First, the Born-geometric structure of modular spacetime, together with the modular uncertainty relation, excludes the phase-space origin and gives rise to a finite modular core, thereby removing physical access to the Schwarzschild singularity. This kinematical feature follows directly from the doubled phase-space structure and is independent of the effective spacetime interpretation adopted in the present work. Second, the metaparticle duality constraint induces an effective conformal deformation of the background geometry and generates an anisotropic effective source that violates the radial Null Energy Condition and the Strong Energy Condition near the modular core. The first mechanism is therefore kinematical and reflects the doubled phase-space structure itself, whereas the second is dynamical and provides a concrete realization of how the focusing assumptions entering the Hawking--Penrose singularity theorems may fail. The resulting nonsingular physical evolution therefore emerges not through the introduction of exotic matter nor through an ad hoc modification of Einstein gravity, but as a consequence of the underlying Born-geometric organization of phase space.

Taken together with our previous thermodynamic study
\cite{Chouha2026MetaparticleThermodynamics}, a coherent
picture now emerges. The earlier work showed that black-hole
evaporation terminates at a finite {\color{red} cold} remnant configuration and
revealed the existence of distinct geometric and dual entropy
contributions. While both sectors reproduce the
Bekenstein--Hawking entropy in the infrared regime, the
geometric contribution decreases throughout the evaporation
process and ultimately vanishes at the remnant endpoint,
whereas the dual contribution remains finite.

The present work provides a geometric explanation of this
behavior. As the classical Schwarzschild singularity is
approached,

\begin{equation}
r\rightarrow0,
\end{equation}

the Born-invariant radius becomes

\begin{equation}
\mathcal R^2
=
r^2+\lambda^4\tilde r^2
\longrightarrow
\lambda^4\tilde r^2,
\end{equation}

so that

\begin{equation}
\mathcal R
\longrightarrow
\lambda^2|\tilde r|.
\end{equation}

Near the modular core the effective geometry therefore becomes
increasingly governed by the dual sector. The persistence of
the dual entropy branch at the remnant endpoint is not an
independent thermodynamic phenomenon but a direct consequence
of the Born-geometric phase-space structure itself. As the
geometric radius is driven toward zero, the modular
uncertainty relation prevents the simultaneous suppression of
the dual coordinate, ensuring that the dual sector remains
active even when the geometric sector becomes subdominant.

From this perspective, the modular remnant should not be
viewed simply as a small black hole. Rather, it represents a
transition between two complementary descriptions of gravity.
In the infrared regime, the geometric sector dominates and
spacetime provides an effective description of gravitational
physics. In the ultraviolet regime, the dual sector becomes
dominant and the modular core appears as the geometric imprint
of an underlying non-geometric phase governed by the dual
degrees of freedom of modular spacetime.

If this interpretation is correct, the resolution of the
Schwarzschild singularity reflects something deeper than the
regularization of a particular spacetime solution. It signals
a transition from a spacetime-based description of gravity to
a fundamentally phase-space-based description in which the
dual degrees of freedom become unavoidable.

Several important directions remain open. Perhaps the most important concerns the formulation of gravity directly on Born geometry. The present work has employed an effective construction in which the Born-invariant coordinate $\mathcal R$ is used to build a Schwarzschild-like geometry and the ordinary Einstein equations are imposed on the resulting effective metric. Consequently, the Born-doubled Schwarzschild solution obtained here should not be regarded as a solution of a fundamental gravitational theory defined on the doubled phase space itself. The existence of a consistent Born-geometric action principle and of the corresponding doubled gravitational field equations remains an open question. It would be particularly interesting to determine whether such a theory admits solutions whose effective low-energy description reproduces the modular core and the metaparticle-induced gravitational response derived in the present work. Another 
particularly
promising avenue concerns the quantum kinematics associated
with modular spacetime. The modular uncertainty relation
suggests a natural extension of conventional generalized
uncertainty principles, and preliminary investigations indicate
that the combined structure of metaparticle dynamics and Born
geometry may lead to a Generalized Extended Uncertainty
Principle (GEUP) unifying ultraviolet and infrared
modifications of quantum mechanics within a common phase-space
framework \cite{ChouhaGEUP}.

\par Another important direction concerns the phenomenology of
modular black holes and the observational signatures of the
underlying Born-geometric phase-space structure. Since the
present construction modifies the interior geometry while
preserving the asymptotic Schwarzschild limit, it is natural
to investigate whether observable imprints may arise through
black-hole thermodynamics and remnant physics
\cite{Barrau2014Remnants,ChenAdler2003Remnants},
the late stages of evaporation
\cite{ChenAdler2003Remnants,CarrMureikaNicolini2015},
quasinormal-mode spectra and gravitational-wave ringdown
signals \cite{BertiCardosoStarinets2009,Berti2025BHSpec},
black-hole shadow observables
\cite{Falcke2000Shadow,EventHorizonTelescope2019,
CunhaHerdeiro2018},
near-horizon imaging tests of strong-field gravity
\cite{Psaltis2018,Johannsen2016},
or the dynamics of accretion flows around compact objects
\cite{NarayanMcClintock2013,YuanNarayan2014}.
Such observables provide a natural framework in which
Born-geometric corrections to classical black-hole solutions
may ultimately be confronted with astrophysical data.
A systematic investigation of these signatures could therefore
provide an observational window into the modular black-hole
scenario and potentially distinguish it from other regular
black-hole and remnant models.

\par Equally intriguing are the cosmological implications of the
present construction. The emergence of a finite modular core and the exclusion of the Schwarzschild singularity from the physically accessible region suggest that the same
Born-geometric mechanism may also operate in homogeneous and
isotropic cosmologies. If the phase-space origin is
inaccessible during gravitational collapse, it is natural to
ask whether the classical Big Bang singularity is similarly
excluded. In this case, the earliest stages of cosmic
evolution may be governed by a finite modular phase rather
than an initial singularity. Extending the present framework
to cosmology may therefore provide a unified description of
black-hole interiors, primordial cosmology, and singularity
resolution within a common Born-geometric framework
\cite{ChouhaModularCosmology}.

More broadly, the picture that emerges is that the singular behavior of classical Schwarzschild spacetime may reflect the limitations of describing gravitational physics solely in terms of spacetime degrees of freedom. In the present framework, the physically relevant arena is a doubled phase space endowed with a Born geometry and a modular uncertainty relation. The exclusion of the phase-space origin suggests that configurations appearing singular in a purely spacetime description need not correspond to physically accessible states of the underlying theory.
Instead, spacetime appears as an effective
description arising from a deeper modular phase-space
structure governed by Born geometry and metaparticle
dynamics. Both black-hole and cosmological singularities may
ultimately originate from attempts to localize physical states
beyond the domain of validity of classical spacetime itself.

The modular core obtained here may therefore represent the
first explicit realization of a more general principle:
the fundamental arena of quantum gravity is not spacetime
alone, but phase space itself.

\appendix
\section{Dimensional Analysis and Scaling} \label{Appendix:A} This appendix provides the dimensional justification for the Born-invariant radial coordinate introduced in Sec.~\ref{Sec: III}. As discussed in Remark~\ref{Rmk:r_phys}, the fundamental metaparticle coordinates satisfy \[ [x^\mu]=L, \qquad [\tilde x_\mu^{\rm phys}]=L. \] In the effective Born-geometric construction developed in Secs.~III and IV, it is convenient to introduce the rescaled dual coordinates \[ \tilde x_\mu = \frac{\tilde x_\mu^{\rm phys}}{\lambda^2}, \] which satisfy \[ [\tilde x_\mu] = L^{-1}. \] Throughout the present appendix, the symbol \(\tilde x_\mu\) refers to these rescaled dual coordinates. The purpose of the rescaling is to allow geometric and dual quantities to contribute symmetrically to Born-invariant distance measures. Since \[ [\lambda]=L, \] the combination \begin{equation} \lambda^2\tilde x_\mu \label{eq:lambda_scaling} \end{equation} has dimensions of length and may therefore be directly compared with the spacetime coordinate \(x^\mu\). \subsection{Most General Quadratic Form} We now construct the simplest Born-geometric line element compatible with \begin{itemize} \item coordinate covariance, \item symmetry between geometric and dual sectors, \item dimensional consistency. \end{itemize} Since \[ [dx^\mu dx^\nu] = L^2, \] and \[ [d\tilde x_\mu d\tilde x_\nu] = L^{-2}, \] the dual contribution must be accompanied by a factor with dimensions \(L^4\). The unique quadratic form satisfying these requirements is therefore \begin{equation} ds^2 = g_{\mu\nu}(x,\tilde x)\, dx^\mu dx^\nu + \lambda^4 g^{\mu\nu}(x,\tilde x)\, d\tilde x_\mu d\tilde x_\nu . \label{eq:born_metric} \end{equation} The factor \(\lambda^4\) ensures that both contributions carry dimensions of length squared, \[ [\lambda^4 d\tilde x_\mu d\tilde x_\nu] = L^4\times L^{-2} = L^2. \] The line element (\ref{eq:born_metric}) is therefore dimensionally consistent and treats the geometric and dual sectors on an equal footing.

\subsection{Flat Phase-Space Limit}

To extract the invariant radial structure, we first consider the simplest case where the metric is locally flat:
\begin{equation}
g_{\mu\nu} = \delta_{\mu\nu}.
\label{eq:flat_metric}
\end{equation}

Substituting \eqref{eq:flat_metric} into \eqref{eq:born_metric}, the line element reduces to
\begin{equation}
ds^2 = dx^i dx^i + \lambda^4\, d\tilde{x}_i d\tilde{x}_i.
\label{eq:flat_phase_space}
\end{equation}

\subsection{Reduction to Spherical Symmetry}

We now introduce spherical coordinates in both sectors.

For the geometric coordinates:
\begin{equation}
r^2 = x^i x^i,
\label{eq:r_definition}
\end{equation}

and similarly for the dual coordinates:
\begin{equation}
\tilde{r}^2 = \tilde{x}_i \tilde{x}_i.
\label{eq:rtilde_definition}
\end{equation}

The differentials satisfy the well-known relations:
\begin{equation}
dx^i dx^i = dr^2 + r^2 d\Omega^2,
\label{eq:dx_spherical}
\end{equation}
and analogously,
\begin{equation}
d\tilde{x}_i d\tilde{x}_i = d\tilde{r}^2 + \tilde{r}^2 d\tilde{\Omega}^2.
\label{eq:dxtilde_spherical}
\end{equation}

\subsection{Restriction to the Radial Sector}

We now focus on purely radial configurations by setting
\begin{equation}
d\Omega = 0, \qquad d\tilde{\Omega} = 0.
\label{eq:radial_restriction}
\end{equation}

Using \eqref{eq:dx_spherical} and \eqref{eq:dxtilde_spherical}, this yields
\begin{equation}
dx^i dx^i = dr^2,
\end{equation}
\begin{equation}
d\tilde{x}_i d\tilde{x}_i = d\tilde{r}^2.
\end{equation}

Substituting into \eqref{eq:flat_phase_space}, we obtain
\begin{equation}
ds^2_{\text{radial}} = dr^2 + \lambda^4 d\tilde{r}^2.
\label{eq:radial_metric}
\end{equation}

\subsection{Invariant Phase-Space Radius}

We now define the invariant phase-space radial distance by integrating \eqref{eq:radial_metric}.

Since $r$ and $\tilde{r}$ are independent variables, we obtain directly
\begin{equation}
\boxed{
\mathcal{R}^2 = r^2 + \lambda^4 \tilde{r}^2
}
\label{eq:R_phase_space}
\end{equation}\\

\subsection{Physical Interpretation}

The quantity $\mathcal{R}$ defined in \eqref{eq:R_phase_space} plays the role of a generalized radial coordinate in phase space.

Unlike the ordinary geometric radius \(r\), the Born-invariant radius \(\mathcal R\) receives contributions from both the geometric and dual sectors. Consequently, the limit \(r\to0\) does not imply \(\mathcal R\to0\) unless the dual coordinate simultaneously vanishes.

This implies that the structure of phase space prevents the simultaneous collapse of the geometric and dual radial sectors and suggests an effective limitation on localization at short distances.
This feature will be central to the resolution of curvature singularities in the effective black hole solutions constructed in this work.
\section*{Acknowledgments}

The author would like to thank Robert Brandenberger for valuable discussions and for his insightful comments on the manuscript.

\bibliographystyle{apsrev4-2}
\bibliography{references-DSBH}

\end{document}